\documentclass[twocolumn,preprintnumbers,a4paper,
superscriptaddress,floatfix,twoside,prl]{revtex4}

\usepackage{bm}
\usepackage{epsfig}
\usepackage{graphics}
\usepackage{natbib}
\usepackage{longtable}
\usepackage{fancyh}
\usepackage[l]{floatflt}
\usepackage{epsfig}
\usepackage{amssymb}
\usepackage{latexsym}
\usepackage{times}
\usepackage{slashed}
\usepackage{upgreek}
\usepackage{amsmath}
\usepackage{amsfonts}
\usepackage{amsbsy}
\usepackage{amscd}
\usepackage{bbm}

\newcommand{\Eref}[1]{Eq.~(\ref{#1})}
\newcommand{\Erefs}[2]{Eqs.~(\ref{#1}) -- (\ref{#2})}

\newcommand{\Fref}[1]{Fig.~\ref{#1}}
\newcommand{\Tref}[1]{Table~\ref{#1}}
\newcommand{\cref}[1]{Ref.~\cite{#1}}
\newcommand{\crefs}[1]{Refs.~\cite{#1}}

\newcommand{\hepph}[1]{{\ftn\tt hep-ph/#1}}

\newcommand{\astroph}[1]{{\ftn\tt astro-ph/#1}}
\newcommand{\arxiv}[1]{{\ftn\tt  arXiv:#1}}

\newcommand{\bal}{\begin{align}}
\newcommand{\eal}{\end{align}}
\newcommand{\beqs}{\begin{subequations}}
\newcommand{\eeqs}{\end{subequations}}
\newcommand{\eec}{\end{center}}
\newcommand{\bec}{\begin{center}}

\newcommand{\eem}{\end{matrix}}
\newcommand{\bem}{\begin{matrix}}
\newcommand{\eeq}{\end{equation}}
\newcommand{\beq}{\begin{equation}}
\newcommand{\ba}{\begin{array}}
\newcommand{\ea}{\end{array}}
\newcommand{\bea}{\begin{eqnarray}}
\newcommand{\eea}{\end{eqnarray}}
\newcommand{\baq}{\begin{eqnarray}}
\newcommand{\eaq}{\end{eqnarray}}

\newcommand\eqs[2]{Eqs.~(\ref{#1}) and (\ref{#2})}

\newcommand{\ftn}{\footnotesize}

\newcommand{\GeV}{{\mbox{\rm GeV}}}

\newcommand{\etal}{{\it et al.\/}}

\def\to{\rightarrow}

\def\llgm{\left\lgroup}
\def\rrgm{\right\rgroup}
\def\lf{\left(}
\def\rg{\right)}
\newcommand\vev[1]{\langle {#1} \rangle}

\newcommand{\Vhi}{\ensuremath{\widehat V_{\rm HI}}}

\newcommand{\Hhi}{\ensuremath{\widehat H_{\rm HI}}}

\newcommand{\Khi}{\ensuremath{K}}

\newcommand{\Vhio}{\ensuremath{\widehat V_{\rm HI0}}}
\newcommand{\Ns}{\ensuremath{{\what N_\star}}}

\newcommand{\mP}{\ensuremath{m_{\rm P}}}
\newcommand{\Mpq}{\ensuremath{M}}

\newcommand{\Mgut}{\ensuremath{M_{\rm GUT}}}

\newcommand{\Ggut}{\ensuremath{G_{\rm GUT}}}

\def\openone{\leavevmode\hbox{\small1\kern-3.8pt\normalsize1}}
\newcommand{\dV}{\ensuremath{\Delta\widehat V_{\rm HI}}}

\newcommand{\fr}{\ensuremath{f_{\cal R}}}

\newcommand{\fm}{\ensuremath{F_{-}}}
\newcommand{\fp}{\ensuremath{F_{+}}}

\newcommand{\hk}{\ensuremath{F_{\rm K}}}
\newcommand{\hr}{\ensuremath{F_{\cal R}}}

\newcommand{\kx}{\ensuremath{k_S}}
\newcommand{\kpp}{\ensuremath{k_\Phi}}
\newcommand{\ksp}{\ensuremath{k_{S\Phi}}}
\newcommand{\ca}{\ensuremath{c_{\cal R}}}

\newcommand{\Gsm}{\ensuremath{G_{\rm SM}}}

\newcommand{\msn}{\ensuremath{\what m_{\rm \dph}}}

\newcommand{\ks}{\ensuremath{k_\star}}
\newcommand{\ns}{\ensuremath{n_{\rm s}}}

\newcommand{\as}{\ensuremath{a_{\rm s}}}
\newcommand{\As}{\ensuremath{A_{\rm s}}}
\newcommand{\rw}{\ensuremath{r_{0.002}}}
\newcommand{\rs}{\ensuremath{r_{\pm}}}

\newcommand{\rce}{\ensuremath{\widehat{\mathcal{R}}}}
\newcommand{\Ve}{\ensuremath{\widehat{V}}}

\newcommand{\dph}{\ensuremath{\delta\phi}}

\newcommand{\phc}{\ensuremath{\Phi}}
\newcommand{\phcb}{\ensuremath{\bar\Phi}}

\newcommand{\what}{\ensuremath{\widehat}}

\def\bbet{{\bar\beta}}
\def\al{{\alpha}}

\def\bt{{\beta}}

\def\n{\bar{n}}

\def\th{{\theta}}
\def\thb{{\bar\theta}}

\def\thn{{\theta_{\Phi}}}

\newcommand{\sg}{\ensuremath{\phi}}

\newcommand{\sgx}{\ensuremath{\phi_\star}}
\newcommand{\sgf}{\ensuremath{\phi_{\rm f}}}

\newcommand{\ld}{\ensuremath{\lambda}}
\newcommand{\ldu}{\ensuremath{\uplambda}}
\newcommand{\Ld}{\ensuremath{\Lambda}}
\newcommand{\kp}{\ensuremath{\kappa}}

\newcommand{\se}{\ensuremath{\widehat \phi}}
\newcommand{\sex}{\ensuremath{\widehat{\phi}_\star}}

\newcommand{\geu}{\ensuremath{\widehat g}}

\newcommand\mtta[4]{\mbox{
$\llgm\bem #1 &#2 \cr #3& #4\eem\rrgm$}}

\def\trns{transplanckian}
\def\Kap{K\"{a}hler potential}

\def\bcp{{\sc\small Bicep2}/{\it Keck Array}}
\newcommand{\plk}{{\it Planck}}

\newcommand{\diag}{\ensuremath{{\sf diag}}}

\newcommand{\cm}{\ensuremath{c_{-}}}
\newcommand{\cp}{\ensuremath{c_{+}}}

\renewcommand{\refname}{{\bf\scshape References}}

\renewenvironment{subequations}{%
\refstepcounter{equation}%
\setcounter{parentequation}{\value{equation}}%
  \setcounter{equation}{0}
  \ignorespaces
}{%
  \setcounter{equation}{\value{parentequation}}%
  \ignorespacesafterend
}

\begin{document}


\title{\bf\scshape Kinetically Modified Non-Minimal Higgs Inflation in Supergravity}

\author{\scshape Constantinos Pallis\\ {\it
Department of Physics, University of Cyprus, P.O. Box 20537,
CY-1678 Nicosia, CYPRUS}\\  {\sl e-mail address: }{\ftn\tt
cpallis@ucy.ac.cy}}



\begin{abstract}

\noindent {\ftn \bf\scshape Abstract:} We consider models of
chaotic inflation driven by the real parts of a conjugate pair of
Higgs superfields involved in the spontaneous breaking of a grand
unification symmetry at a scale assuming its supersymmetric value.
We combine a superpotential, which is uniquely determined by
applying a continuous $R$ symmetry, with a class of logarithmic or
semi-logarithmic \Kap s which exhibit a prominent shift-symmetry
with a tiny violation, whose strengths are quantified by $\cm$ and
$\cp$ respectively. The inflationary observables provide an
excellent match to the recent \bcp\ and \plk\ results setting
$3.5\cdot10^{-3}\lesssim\rs=\cp/\cm\lesssim1/N$ where $N=3$ or $2$
is the prefactor of the logarithm. Inflation can be attained for
subplanckian inflaton values with the corresponding effective
theories retaining the perturbative unitarity up to the Planck
scale.
\\ \\ {\scriptsize {\sf PACs numbers: 98.80.Cq, 04.50.Kd, 12.60.Jv, 04.65.+e}
\hfill {\sl\bfseries Published in} {\sl Phys. Rev. D} {\bf 92},
no. 12, 121305(R) (2015)}

\end{abstract}\pagestyle{fancyplain}

\maketitle

\rhead[\fancyplain{}{ \bf \thepage}]{\fancyplain{}{\sl Kinetically
Modified nMHI in SUGRA}} \lhead[\fancyplain{}{\sl C.
Pallis}]{\fancyplain{}{\bf \thepage}} \cfoot{}

\section{Introduction}

Soon after inflation's \cite{guth} introduction as a solution to a
number of longstanding cosmological puzzles -- such as the horizon
and flatness problems -- many efforts have been made so as to
connect it with a \emph{Grand Unified Theory} ({\sf \ftn GUT})
phase transition in the early universe -- see e.g. \crefs{old,
fhi, jones2, nmH, lazarides, fhi1,fhi2,fhi3,okada}. According to
this economical and highly appealing set-up, the scalar field
which drives inflation (called inflaton) plays, at the end of its
inflationary evolution, the role of a Higgs field
\cite{old,jones2,nmH,okada, lazarides} or destabilizes others
fields, which act as Higgs fields \cite{fhi,fhi1,fhi2,fhi3,fhi4}.
As a consequence, a GUT gauge group $\Ggut$ can be spontaneously
broken after the end of inflation. The first mechanism above can
be also applied in the context of the \emph{Standard Model}
({\sf\ftn SM}) \cite{sm} or the next-to-\emph{Minimal
Supersymmetric SM} ({\sf\ftn MSSM}) \cite{linde1, shiftHI} and
leads to the spontaneous breaking of the electroweak gauge group
${G_{\rm SM}}= SU(3)_{\rm C}\times SU(2)_{\rm L}\times U(1)_{Y}$
by the Higgs/inflaton field(s).

We here focus on the earlier version of this idea -- i.e. the
GUT-scale Higgs inflation -- concentrating on its
\emph{supersymmetric} ({\sf \ftn SUSY}) realization \cite{fhi,
jones2, nmH, lazarides, fhi1,fhi2,fhi3,fhi4, okada}, where the
notorious GUT hierarchy problem is elegantly addressed. The
starting point of our approach is the simplest superpotential
\beq W=\ld S\lf\bar\Phi\Phi-\Mpq^2/4\rg\label{Win} \eeq
which leads to the spontaneous breaking of $\Ggut$ and is uniquely
determined, at renormalizable level, by a convenient \cite{fhi}
continuous $R$ symmetry. Here, $\ld$ and $M$ are two constants
which can both be taken positive by field redefinitions; $S$ is a
left-handed superfield, singlet under $\Ggut$; $\bar\Phi$ and
$\Phi$ is a pair of left-handed superfields belonging to
non-trivial conjugate representations of $\Ggut$, and reducing its
rank by their \emph{vacuum expectation values} ({\sf\ftn v.e.vs})
-- see e.g. \crefs{fhi1, fhi2}. Just for definiteness we restrict
ourselves to $\Ggut=G_{\rm SM}\times U(1)_{B-L}$
\cite{fhi1,okada}, gauge group which consists the simplest GUT
beyond the MSSM -- where $B$ and $L$ denote the baryon and lepton
number. With the specific choice of $\Ggut$ $\Phi$ and $\bar\Phi$
carry $B-L$ charges $1$ and $-1$ respectively.

Moreover, $W$ combined with a judiciously selected \Kap, $K$,
gives rise to two types of inflation, in the context of
\emph{Supergravity} ({\sf\ftn SUGRA}). In particular, we can
obtain \emph{F-term hybrid inflation} ({\sf \ftn FHI}) driven by
$S$ with $\bar\Phi$ and $\Phi$ being confined to zero or
\emph{non-minimal Higgs inflation} ({\sf \ftn nMHI}),
interchanging the roles of $S$ and $\bar\Phi-\Phi$. A canonical
\cite{fhi1} or quasi-canonical \cite{fhi3, fhi2} $K$ is convenient
for implementing FHI, whereas a logarithmic $K$ including an
holomorphic function $\ca\phc\phcb$ with large $\ca>0$ \cite{nmH}
or tiny $\ca<0$ \cite{okada} is dictated for nMHI. Although FHI
can become compatible with data \cite{plcp} at the cost of a mild
tuning of one \cite{fhi1,fhi3} (or more \cite{fhi2}) parameters
beyond $\ld$ and $M$, it exhibits a serious drawback which can be
eluded, by construction, in nMHI. Since $\Ggut$ is broken only at
the SUSY vacuum, after the end of FHI, topological defects are
formed, if they are predicted by the $\Ggut$ breaking. This does
not occur within nMHI since $\Ggut$ is already spontaneously
broken during it, through the non-zero $\phcb$ and $\phc$ values.
Utilizing large enough $\ca$'s \cite{nmH} or adjusting three
parameters ($\ld$, $\ca$ and $M$) \cite{okada}, acceptable values
for the (scalar) spectral index, $\ns$, can be achieved with low
enough \cite{nmH} or higher \cite{okada} tensor-to-scalar ratio,
$r$. In the former case, though, the largeness of $\ca$ violates
the perturbative unitarity \cite{cutoff, riotto} whereas in the
latter case, \trns\ values of the inflaton jeopardize the validity
of the inflationary predictions.

In this letter, we show that the shortcomings above can be
elegantly overcome, if we realize the recently proposed
\cite{nMkin} idea of \emph{kinetically modified non-minimal
inflation} with a $\Ggut$ non-singlet inflaton. The crucial
difference of this setting compared to the nMHI with large $\ca$
\cite{nmH} is that the slope of the inflationary potential and the
canonical normalization of the higgs-inflaton do not depend
exclusively on one parameter, $\ca$, but separately on two
parameters, $\cp$ and $\cm$, whose the ratio $\rs=\cp/\cm\ll1$
determines $\ns$ and $r$. In particular, restricting $\rs$ to
natural values, motivated by an enhanced shift symmetry, the
inflationary observables can nicely cover the 1-$\sigma$ domain of
the present data \cite{plcp,gws},
\beq \label{data} \ns=0.968\pm0.0045~~\mbox{and}~~r=
0.048^{+0.035}_{-0.032}, \eeq
independently of $M$ which may be confined precisely at its value
entailed by the gauge unification within MSSM. Contrary to our
recent investigation \cite{lazarides}, where we stick to quadratic
terms for $\Phi$ and $\bar\Phi$ in the selected $K$'s we here
parameterize the relevant terms with an exponent $m$. Moreover, we
here, insist to integer prefactors of the logarithms involved in
$K$'s, increasing thereby the naturalness of the model. As regards
other simple and well-motivated inflationary models
\cite{roest,eno5} which share similar inflationary potentials with
the one obtained here, let us underline that the use of a gauge
non-singlet inflaton with subplanckian values together with the
enhanced resulting $r$'s, in accordance with an approximate shift
symmetry, consist the main novelties of our approach.

Below we describe a class of \Kap s which lead to kinetically
modified nMHI, we outline the derivation of the inflationary
potential and restrict the free parameters of the models testing
them against observations. Finally, we analyze the
\emph{ultraviolet} ({\sf\ftn UV}) behavior of these models and
summarize our conclusions.

\section{K\"{a}hler Potentials} The key ingredient
of our proposal is the selection of a purely or partially
logarithmic $K$ including the real functions
\beq \label{fs}
F_\pm=\left|\Phi\pm\bar\Phi^*\right|^2~~\mbox{and}~~F_S=|S|^2-\kx{|S|^4},
\eeq
which respect the symmetries of $W$ -- star ($^*$) denotes complex
conjugation. As we show below, $\cm\fm$ dominates the canonical
normalization of inflaton, $\cp\fp$ plays the role of the
non-minimal inflaton-curvature coupling and $F_S$ provides a
typical kinetic term for $S$, considering the next-to-minimal term
for stability/heaviness reasons \cite{linde1}. Obviously, $F_S$ is
the same as that used in \cref{nMkin}, apart from an overall
normalization factor, whereas $\fm$ and $\fp$ correspond to $\hk$
and $\hr$ respectively. However, $\fp$ is a real and not an
holomorphic function as $\hr$. Actually, it remains invariant
under the transformation $\Phi\ \to\ \Phi+ c$ and $\bar\Phi\ \to\
\bar\Phi-c^*$ (where $c$ is a complex number) whereas $\fm$
respects the symmetry $\Phi\ \to\ \Phi+ c$ and $\bar\Phi\ \to\
\bar\Phi+c^*$ which coincides with the former only for $c=0$.
Stability of the selected inflationary direction entails that the
latter symmetry is to be the dominant one -- see below. The
particular importance of the shift symmetry in taming the
so-called $\eta$-problem of inflation in SUGRA is first recognized
for gauge singlets in \cref{kawasaki} and non-singlets in
\cref{shiftHI}.


%

In terms of the functions introduced in \Eref{fs} we postulate the
following form of $K$
\beqs\bea \nonumber  K_1&=&-3\ln\left(1+\cp\fp-\frac13(1+\cp\fp)^m\cm\fm \right.\\
&& \left.-\frac13F_S+\kpp\fm^2+
\frac13\ksp\fm{|S|^2}\right)\,,~~~~~\label{K1}\eea
where we take for consistency all the possible terms up to fourth
order whereas a term of the form $-k_{S+}\fp^m |S|^2/3$ is
neglected for simplicity, given that $\fp$ is considered as a
violation of the principal symmetry -- we use throughout units
with the reduced Planck scale $\mP = 2.433\cdot 10^{18}~\GeV$
being set equal to unity. Identical results can be achieved if we
select $K=K_2$ with
\beq
K_2=-3\ln\left(1+\cp\fp-\frac13F_S\right)+\frac{\cm\fm}{(1+\cp\fp)^{1-m}}\,\cdot\label{K2}\eeq
If we place $F_S$ outside the argument of the logarithm, we can
obtain two other $K$'s -- not mentioned in \cref{nMkin} -- which
lead to similar results. Namely, \beq
K_3=-2\ln\lf1+\cp\fp-\frac12(1+\cp\fp)^m\cm\fm\rg+F_S\,\label{K3}\eeq
and \beq
K_4=-2\ln(1+\cp\fp)+F_S+(1+\cp\fp)^{m-1}\cm\fm\,.\label{K4}\eeq\eeqs
To highlight the robustness of our setting we use only integer
prefactors for the logarithms avoiding thereby any relevant tuning
-- cf.~\cref{lazarides, nIG}. Note that for $m=0$ [$m=1$], $\fm$
and $\fp$ in $K_1$ and $K_3$ [$K_2$ and $K_4$] are totally
decoupled, i.e. no higher order term is needed. If we allow for a
continuous variation of the $\ln$ prefactor, too, we can obtain
several variants of kinetically modified nMHI. For $m=0$ this
possibility is analyzed in \cref{lazarides}.

Given that $M\ll1$ does not affect the inflationary epoch, the
free parameters of our models, for fixed $m$, are $\rs$ and
$\ld/\cm$ and not $\cm$, $\cp$ and $\ld$ as naively expected.
Indeed, performing the rescalings $\Phi\to\Phi/\sqrt{\cm}$ and
$\bar\Phi\to\bar\Phi/\sqrt{\cm}$, in Eqs.~(\ref{Win}) and
(\ref{K1}) -- (\ref{K4}) we see that $W$ and $K$ depends
exclusively on $\ld/\cm$ and $\rs$ respectively. Therefore, our
models are equally economical as nMHI with $\ca<0$ \cite{okada}
and they have just one more free parameter than nMHI with $\ca>0$
\cite{nmH} -- see also \cref{roest}. Unlike these models, however,
-- where the largeness \cite{nmH} or the smallness \cite{okada} of
$\ca$ can not be justified by any symmetry -- our models can be
characterized as completely natural, in the 't Hooft sense, since
in the limits $\rs=\cp/\cm\to0$ and $\ld\to0$, they enjoy the
following enhanced symmetries:
\beq \Phi \to\ \Phi+c,\>\>\>\bar\Phi \to\ \bar\Phi+c^*
\>\>\>\mbox{and}\>\>\> S \to\ e^{i\varphi} S,\label{shift}\eeq
where $c$ and $\varphi$ is a complex and a real number
respectively. The same argument guarantees the smallness of
$k_{S+}$ in a possible term  $-k_{S+}\fp^m|S|^2/3$ inside the
logarithms in \Eref{K1} or \Eref{K2}. On the other hand, our
models do not exhibit any no-scale-type symmetry like that
postulated in \cref{eno5}.

\renewcommand{\arraystretch}{1.2}
\begin{table*}[!t]
\caption{\normalfont Mass-squared spectrum for $K=K_i$ and
$K=K_{i+2}$ ($i=1,2$) along the path in \Eref{inftr}.}
\begin{tabular}{c@{\hspace{0.5cm}}|c|c@{\hspace{0.5cm}}|c@{\hspace{0.5cm}}|
c@{\hspace{0.5cm}}|c@{\hspace{0.5cm}}}\toprule
%
{\sc Fields}&{\sc Eigenstates} & \multicolumn{4}{c}{\sc Masses
Squared}\\\cline{3-6}
&&\hspace*{0.5cm}{\sc Symbol} & \hspace*{0.5cm}{$K=K_1$}&\hspace*{0.5cm}{$K=K_2$} & \hspace*{0.5cm}{$K=K_{i+2}$} \\
\colrule
%
%
2 real scalars&$\widehat\theta_{+}$&\hspace*{0.5cm} $\widehat
m_{\theta+}^2$& \hspace*{0.5cm} $4\Hhi^2$
&\multicolumn{2}{|c}{$6\Hhi^2$}\\
&$\widehat \theta_\Phi$ &\hspace*{0.5cm} $\widehat m_{
\theta_\Phi}^2$& \hspace*{0.5cm} $M^2_{BL}+4\Hhi^2$&
\multicolumn{2}{|c}{$M^2_{BL}+6\Hhi^2$}\\\cline{4-6}
1 complex scalar&$\widehat s, \widehat{\bar{s}}$ &\hspace*{0.5cm}
$ \widehat m_{
s}^2$&\multicolumn{2}{c|}{$6\lf2\kx\fr-1/3\rg\Hhi^2$}&$12\kx\Hhi^2$\\\hline
1 gauge  boson& $A_{BL}$ &
\hspace*{0.5cm} $ M_{BL}^2$&\multicolumn{3}{c}{$g^2\cm\lf\fr^{m-1}-N\rs /\fr\rg\sg^2$}\\
\colrule
$4$ Weyl spinors & $\what \psi_\pm
=\frac{1}{\sqrt{2}}(\what{\psi}_{\Phi+}\pm \what{\psi}_{S})$ &
\hspace*{0.5cm} $\what m^2_{ \psi\pm}$ &
\multicolumn{3}{c}{\hspace*{0.5cm}${24\Hhi^2/\cm\sg^2\fr^{1+m}}$}
\\\cline{4-6} &$\ldu_{BL}, \widehat\psi_{\Phi-}$&\hspace*{0.5cm}
$M_{BL}^2$&\multicolumn{3}{c}{$g^2\cm\lf\fr^{m-1}-N\rs
/\fr\rg\sg^2$}\\\botrule
\end{tabular}\label{tab1}
\end{table*}
%

\section{Inflationary Potential}

The \emph{Einstein frame} ({\sf\ftn EF}) action within SUGRA for
the complex scalar fields $z^\al=S,\phc,\phcb$ -- denoted by the
same superfield symbol -- can be written as \cite{linde1}
\beqs\beq\label{Saction1}  {\sf S}=\int d^4x \sqrt{-\what{
\mathfrak{g}}}\lf-\frac{1}{2}\rce +K_{\al\bbet} \geu^{\mu\nu}D_\mu
z^\al D_\nu z^{*\bbet}-\Ve\rg \eeq
where summation is taken over $z^\al$; $\rce$ is the EF Ricci
scalar curvature; $D_\mu$ is the gauge covariant derivative,
$K_{\al\bbet}=K_{,z^\al z^{*\bbet}}$ and
$K^{\al\bbet}K_{\bbet\gamma}=\delta^\al_{\gamma}$ -- the symbol
$,z$ as subscript denotes derivation \emph{with respect to}
({\ftn\sf w.r.t}) $z$. Also $\Ve$ is the EF SUGRA potential which
can be found in terms of $W$ in \Eref{Win} and the $K$'s in
Eqs.~(\ref{K1}) -- (\ref{K4}) via the formula
\beq \hspace*{-1.5mm}\Ve=e^{\Khi}\left(K^{\al\bbet}D_\al
WD^*_\bbet W^*-3{\vert W\vert^2}\right)+\frac{g^2}2
\mbox{$\sum_a$} {\rm D}_a {\rm D}_a,\label{Vsugra} \eeq\eeqs
where $D_\al W=W_{,z^\al} +K_{,z^\al}W$, ${\rm D}_a=z_\al\lf
T_a\rg^\al_\bt K^\bt$ and the summation is applied over the
generators $T_a$ of $\Ggut$. If we express $\Phi, \bar\Phi$ and
$S$ according to the parametrization
\beq\label{hpar} \Phi=\frac{\sg e^{i\th}}{\sqrt{2}}
\cos\thn,~~\bar\Phi=\frac{\sg e^{i\thb}}{\sqrt{2}}
\sin\thn,~~\mbox{and}~~S=\frac{s +i\bar s}{\sqrt{2}}\,,\eeq
with $0\leq\thn\leq\pi/2$, we can easily deduce from \Eref{Vsugra}
that a D-flat direction occurs at
\beq \label{inftr} \bar
s=s=\th=\thb=0\>\>\>\mbox{and}\>\>\>\thn={\pi/4}\eeq
along which the only surviving term in \Eref{Vsugra} is
\beqs\beq \label{Vhio}\Vhi=e^{K}K^{SS^*}\,
|W_{,S}|^2\,=\frac{\ld^2(\sg^2-M^2)^2}{16\fr^{2}}\,,\eeq
since we obtain
\beq \label{Nab} K^{SS^*}=\begin{cases} \fr\\1\end{cases}
\hspace*{-0.2cm}\mbox{for}~~ K=\begin{cases} K_i\\ K_{i+2}
\end{cases}\mbox{with}~~i=1,2 \eeq\eeqs
where $\fr=1+\cp\sg^2$ plays the role of a non-minimal coupling to
Ricci scalar in the \emph{Jordan frame} ({\sf\ftn JF}). Indeed, if
we perform a conformal transformation \cite{linde1, lazarides,
nIG} defining the frame function as
${\Omega/N}=-\exp\lf-{K}/{N}\rg$, where
\beq
N=3~~\mbox{or}~~N=2~~\mbox{for}~~K=K_i~~\mbox{or}~~K=K_{i+2}\,,
\eeq
respectively, we can easily show that $\fr=-\Omega/N$ along the
path in \Eref{inftr}. It is remarkable that $\Vhi$ turns out to be
independent of the coefficients $\cm, \kpp$ and $\ksp$ in
Eqs.~(\ref{K1}) -- (\ref{K2}). Had we introduced the term
$-k_{S+}\fp^m|S|^2/3$ inside the logarithms in \eqs{K1}{K2}, we
would have obtained an extra factor $\lf1+k_{S+}\sg^{2m}\rg$ in
the denominator of $\Vhi$. Our results remain intact from this
factor provided that $k_{S+}\leq0.001$. Note, finally, that the
conventional Einstein gravity is recovered at the SUSY vacuum,
\beq \label{vevs} \vev{S}=0~~~\mbox{and}~~~\vev{\phi}=M\ll1\eeq
since $\vev{\fr}\simeq1$.

To specify the EF canonically normalized inflaton, we note that,
for all choices of $K$ in \Erefs{K1}{K4}, $K_{\al\bbet}$ along the
configuration in \Eref{inftr} takes the form
\beq \lf K_{\al\bbet}\rg=\diag\lf M_K,K_{SS^*}\rg~~\mbox{with}~~
M_K=\frac{1}{\fr^2}\mtta{\kappa}{\bar\kappa}{\bar\kappa}{\kappa},\label{Sni1}
\eeq
where $\kp=\cm\fr^{1+m}-N\cp$ and $\bar\kp={N\cp^2\sg^2}$.  Upon
diagonalization of $M_K$ we find its eigenvalues which are
\beqs\bea
\label{kp}\kp_+&=&\cm\lf\fr^{1+m}+N\rs(\cp\sg^2-1)\rg/{\fr^2};\\
\label{km} \kp_-&=&\cm\lf\fr^m- {N\rs}\rg/{\fr},~~~\eea\eeqs
where the positivity of $\kp_-$ is assured during and after nMHI
for $\rs\lesssim1/N$ given that $\vev{\fr}\simeq1$. Inserting
\eqs{hpar}{Sni1} in the second term of the \emph{right-hand side}
({\sf\ftn r.h.s}) of \Eref{Saction1} we can define the EF
canonically normalized fields which are denoted by hat and are
found to be
\beqs\bea \label{VJe} &&
\frac{d\se}{d\sg}=J=\sqrt{\kp_+},\>\>\widehat{\theta}_+
={J\sg\theta_+\over\sqrt{2}}, \widehat{\theta}_-
=\sqrt{\frac{\kp_-}{2}}\sg\theta_-\,,~~~~~~~~~~~~\\ && \widehat
\theta_\Phi = \sg\sqrt{\kp_-}\lf\theta_\Phi-{\pi/4}\rg,~~(\what
s,\what{\bar{s}})=\sqrt{K_{SS^*}}(s,\bar s)\,,~~~\eea\eeqs
where $\th_{\pm}=\lf\bar\th\pm\th\rg/\sqrt{2}$. Note, in passing,
that the spinors $\psi_S$ and $\psi_{\Phi\pm}$ associated with the
superfields $S$ and $\Phi-\bar\Phi$ are normalized similarly,
i.e., $\what\psi_{S}=\sqrt{K_{SS^*}}\psi_{S}$ and
$\what\psi_{\Phi\pm}=\sqrt{\kp_\pm}\psi_{\Phi\pm}$ with
$\psi_{\Phi\pm}=(\psi_\Phi\pm\psi_{\bar\Phi})/\sqrt{2}$.

Taking the limit $\cm\gg\cp$ we find the expressions of the masses
squared $\what m^2_{\chi^\al}$ (with
$\chi^\al=\theta_+,\theta_\Phi$ and $S$) arranged in \Tref{tab1},
which approach rather well the quite lengthy, exact expressions
taken into account in our numerical computation. These expressions
assist us to appreciate the role of $\kx>0$ in retaining positive
$\what m^2_{s}$ for $K=K_i$ and heavy enough for $K=K_{i+2}$.
Indeed, $\what m^2_{\chi^\al}\gg\Hhi^2=\Vhio/3$ for
$\sgf\leq\sg\leq\sgx$ -- where $\sgx$ and $\sgf$ are the values of
$\sg$ when $\ks=0.05/{\rm Mpc}$ crosses the horizon of nMHI and at
its end correspondingly. In \Tref{tab1} we display also the
masses, $M_{BL}$, of the gauge boson $A_{BL}$ -- which signals the
fact that $\Ggut$ is broken during nMHI -- and the masses of the
corresponding fermions.

The derived mass spectrum can be employed in order to find the
one-loop radiative corrections, $\dV$, to $\Vhi$. Considering
SUGRA as an effective theory with cutoff scale equal to $\mP$, the
well-known Coleman-Weinberg formula can be employed
self-consistently taking into account only the masses which lie
well below $\mP$, i.e., all the masses arranged in \Tref{tab1}
besides $M_{BL}$ and $\what m_{\th_\Phi}$. The resulting $\dV$
lets intact our inflationary outputs, provided that the
renormalization-group mass scale $\Lambda$, is determined by
requiring $\dV(\sgx)=0$ or $\dV(\sgf)=0$. The possible dependence
of our findings on the choice of $\Lambda$ can be totally avoided
if we confine ourselves to $\ksp\sim1$ and $\kx\sim1$ resulting to
$\Ld\simeq3.2\cdot10^{-5}-1.4\cdot10^{-4}$. Under these
circumstances, our inflationary predictions can be exclusively
reproduced by using $\Vhi$ in \Eref{Vhio} -- cf. \cref{lazarides}.

\section{Inflationary Requirements}

Applying the standard formulas quoted in \cref{nMkin} for $\what
V_{\rm CI}=\Vhi$, we can compute a number of observational
quantities, which assist us to qualify our inflationary setting.
Namely, we extract the number, $\Ns$, of e-foldings that the scale
$\ks$ experiences during nMHI and the amplitude, $\As$, of the
power spectrum of the curvature perturbations generated by $\sg$
for $\sg=\sgx$. These observables must be compatible with the
requirements \cite{plcp}
\begin{equation}
\label{Ntot}
\Ns\simeq61.5+\ln{\Vhi(\sgx)^{\frac12}\over\Vhi(\sgf)^{\frac14}}~~\mbox{and}~~\As^{\frac12}\simeq4.627\cdot10^{-5},\eeq
where we consider an equation-of-state parameter $w_{\rm int}=1/3$
correspoding to quatric potential which is expected to approximate
rather well $\Vhi$ for $\sg\ll1$. We can, then, compute the model
predictions as regards $\ns$, its running, $\as$ and $r$ or $\rw$
-- see \cref{nMkin}. The analytic expressions displayed in
\cref{nMkin} for these quantities are applicable to our present
case too, for $m>-1$, performing the following replacements:
\beq \label{nmkin} n=4,~r_{\cal R\rm K}=\rs,~~\mbox{and}~~c_{\rm
K}=\cm\,\eeq
and multiplying by a factor of two the r.h.s of the equation which
yields $\ld$ in terms of $\cm$. We here concetrate on $m>-1$ since
for smaller $m$'s, confining  $\ns$ to its allowed region in
\Eref{data} the predicted $r$'s, although acceptable, lie well
below the sencitivity of the present experiments \cite{cmbpol}.
This happens because, decreasing $m$ below $0$, the first term in
the r.h.s of \Eref{kp} becomes progressively subdominant and thus,
$\cp$ controls both the slope of $\Vhi$ and the value of $J$ in
\Eref{VJe} as in the standard nMHI \cite{nmH,okada}.

The inflationary observables are not affected by $M$, provided
that it is confined to values much lower than $\mP$. This can be
done if we determine it identifying the unification scale (as
defined by the gauge-coupling unification within the MSSM)
$\Mgut\simeq2/2.433\cdot10^{-2}$ with the value of $M_{BL}$ -- see
\Tref{tab1} -- at the SUSY vacuum. Given that $\vev{\fr}\simeq1$
and $\vev{\kp_+}\simeq1-{N\rs}$, we obtain, for $\rs\lesssim1/N$,
\beq \label{Mg} M\simeq{\Mgut}/{g\sqrt{\cm\lf1-{N\rs}\rg}}\eeq
with $g\simeq0.7$ being the value of the GUT gauge coupling
constant. This result influences the inflaton mass at the vacuum,
which is estimated to be $\msn\simeq\ld
M/\sqrt{2\cm\lf1-{N\rs}\rg}$.

\begin{figure}[!t]
\includegraphics[width=60mm,angle=-90]{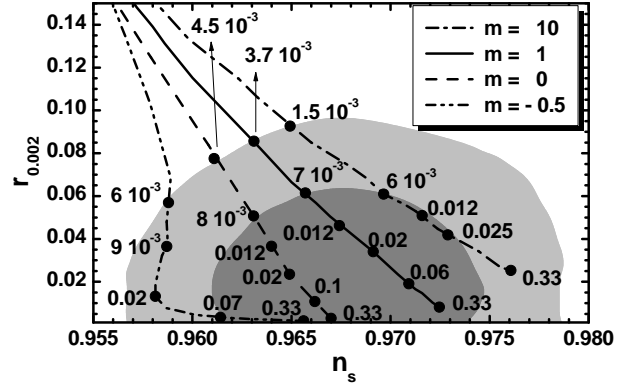}
\caption{\sl Allowed curves in the $\ns-\rw$ plane for $K=K_i$
($i=1,2$) and various $m$'s (shown in the plot legend) and $\rs$'s
indicated on the curves. The marginalized joint $68\%$ [$95\%$]
regions from Planck, \bcp\ and Baryon Acoustic Oscillations (BAO)
data are depicted by the dark [light] shaded
contours.}\label{fig1}
\end{figure}

\section{Results}

Imposing the conditions in \Eref{Ntot}  we restrict $\ld/\cm$ and
$\sex$ whereas \Eref{data} constrains mainly $m$ and $\rs$.
Focusing  initially on $K=K_i$ with $i=1,2$ we present our results
in Figs.~\ref{fig1} and \ref{fig3}. Namely, in \Fref{fig1} we
compare the allowed curves in the $\ns-\rw$ plane with the
observational data \cite{plcp} for $m=-1/2,0,1$ and $10$ -- double
dot-dashed, dashed, solid, and dot-dashed line respectively. The
variation of $\rs$ is shown along each line. Note that for $m=0$
the line essentially coincides with the corresponding one in
\cref{lazarides} -- cf. \crefs{roest,nMkin} -- and declines from
the central $\ns$ value in \Eref{data}. On the other hand, the
compatibility of the $m=1$ line with the central values in
\Eref{data} is certainly impressive. For low enough $\rs$'s --
i.e. $\rs\leq10^{-4}$ -- the various lines converge to the
$(\ns,\rw)$'s obtained within quartic inflation whereas, for
larger $\rs$, they enter the observationally allowed regions and
terminate for $\rs\simeq1/3$, beyond which $\kp_-$ in \Eref{km}
ceases to be well defined. Notably, this restriction provides a
lower bound on $\rw$ which increases with $m$. Indeed, we obtain
$\rw\gtrsim0.0017,0.0028, 0.009$ and $0.025$ for $m=-1/2,0,1$ and
$10$ correspondingly. Therefore, our results are testable in
forthcoming experiments \cite{cmbpol}.


Repeating the same analysis for $(-1)\leq m\leq10$ we can identify
the allowed range of $\rs$ -- as in \Fref{fig3}. The allowed
(shaded) region is bounded by the dashed line, which corresponds
to $\rs\simeq1/3$, and the dot-dashed and thin lines along which
the lower and upper bounds on $\ns$ and $r$ in \Eref{data} are
saturated respectively. We remark that increasing $\rs$, with
fixed $m$, $\ns$ increases whereas $r$ decreases, in accordance
with our findings in \Fref{fig1}. We also infer that $\rs$ takes
more natural (lower than unity) values for larger $m$'s. Fixing
$\ns$ to its central value in \Eref{data} we obtain the solid line
along which we get clear predictions for $r$, $\as$ and $\msn$.
Namely,
\beqs\bea\label{res1} && 0.18\lesssim
m\lesssim10\>\>\>\mbox{and}\>\>\> 1/3\gtrsim
{\rs}\gtrsim3.5\cdot10^{-3};\\ && \label{res2} 0.4\lesssim
{r/0.01}\lesssim7.6\>\>\>\mbox{and}\>\>\>5.4\lesssim-\as/10^{-4}\lesssim6
~~~~~~~~~\eea\eeqs
with $2.4\cdot10^{-8}\lesssim\msn\lesssim8.7\cdot10^{-6}$. Since
the resulting $|\as|$ remains sufficiently low, our models are
consistent with the fitting of data with the $\Lambda$CDM+$r$
model \cite{plcp}. Finally, the $\msn$ range lets open the
possibility of non-thermal leptogenesis \cite{ntlepto} if we
introduce a suitable coupling between $\bar\Phi$ and the
right-handed neutrinos -- see e.g. \crefs{fhi1,nmH}.

\begin{figure}[!t]
\includegraphics[width=60mm,angle=-90]{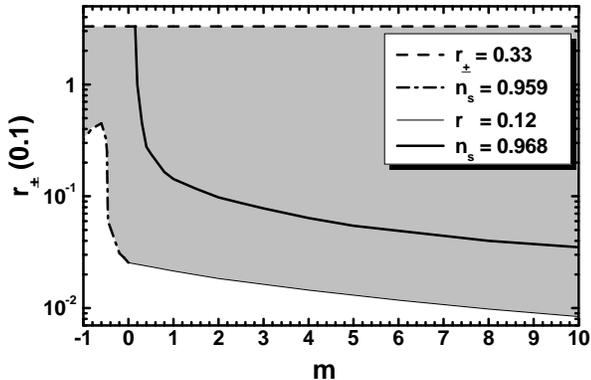}
\caption{\sl Allowed (shaded) region in the $m-\rs$ plane for
$K=K_{i}$. The conventions adopted for the various lines are also
shown.}\label{fig3}
\end{figure}

Had we employed $K=K_{i+2}$, the various lines in \Fref{fig1} and
the allowed regions in Fig.~\ref{fig3} would have been extended
until $\rs\simeq1/2$. This bound would have yielded
$\rw\gtrsim0.0012, 0.002, 0.0066$ and $0.023$ for $m=-1/2,0,1$ and
$10$ correspondingly, which are a little lower than those designed
in \Fref{fig1}. The lower bounds of $m$, $\rs$ and $r$ in
\eqs{res1}{res2} become $0.19$, $1/2$, and $0.003$, the upper
bound on $\msn$ moves on to  $1.3\cdot10^{-5}$ whereas the bounds
on $(-\as)$ remain unaltered.

\begin{figure}[!t]
\includegraphics[width=60mm,angle=-90]{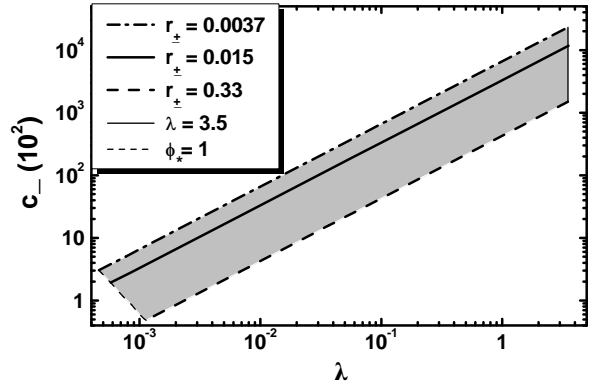}
\caption{\sl Allowed (shaded) region in the $\ld-\cm$ plane for
$K=K_{i}$ (with $i=1,2$) and $m=1$. The conventions adopted for
the various lines are also shown.}\label{fig2}
\end{figure}

Although $\ld/\cm$ is constant in our setting for fixed $\rs$ and
$m$, the amplitudes of $\ld$ and $\cm$ can be bounded. This fact
is illustrated in \Fref{fig2} where we display  the allowed
(shaded) area in the $\ld-\cm$ plane focusing on the $m=1$ case.
We observe that for any $\rs$ between its minimal ($0.0037$) and
maximal ($1/3$) value -- depicted by bold dot-dashed and dashed
lines -- there is a lower bound -- represented by a faint dashed
line -- on $\cm$, above which $\sgx<1$. Consequently, our proposal
can be stabilized against corrections from higher order terms --
e.g., $(\bar\Phi\Phi)^l$ with $l>1$ in \Eref{Win}. The
perturbative bound $\ld=3.5$ limits the region at the other end
along the thin solid line. Plotted is also the solid line for
$\rs=0.015$ which yields $\ns=0.968$. The corresponding $r=0.043$
turns out to be impressively close to its central value in
\Eref{data}.

\section{The Effective Cut-Off Scale}\label{uv}

The fact that $\se$ in \Eref{VJe} does not coincide with $\sg$ at
the vacuum of the theory -- contrary to the pure nMHI
\cite{cutoff, riotto} -- assures that the corresponding effective
theories respect perturbative unitarity up to $\mP=1$ although
$\cm$ may take relatively large values for $\sgx<1$ -- see
\Fref{fig2}. To clarify further this point, we analyze the
small-field behavior of our models in the EF for $m=1$. We focus
on the second term in the r.h.s of \Eref{Saction1} for $\mu=\nu=0$
and we expand it about $\vev{\phi}=M\ll1$ in terms of $\se$. Our
result is written as
\beqs\beq J^2
\dot\phi^2\simeq\lf1+3N\rs^2\se^2-5N\rs^3\se^4+\cdots\rg\dot\se^2.\eeq
Expanding similarly $\Vhi$, see \Eref{Vhio}, in terms of $\se$ we
have
\beq
\Vhi\simeq\frac{\ld^2\what{\sg}^4}{16\cm^{2}}\lf1-2\rs\what{\sg}^{2}+3\rs^2\what{\sg}^4-\cdots\rg.
\eeq\eeqs
Similar expressions can be obtained for the other $m$'s too. Given
that the positivity of $\kp_-$ in \Eref{kp} entails
$\rs\lesssim1/N<1$, we can conclude that our models do not face
any problem with the perturbative unitarity up to $\mP$.


\section{Conclusions and Perspectives}

The feasibility of inflating with a superheavy Higgs field is
certainly an archetypal open question. We here outlined a fresh
look, identifying a class of \Kap s in Eqs.~(\ref{K1}) --
(\ref{K4}) which can cooperate with the superpotential in
\Eref{Win} and lead to the SUGRA potential $\Vhi$ collectively
given by \Eref{Vhio}. Prominent in the proposed \Kap s is the role
of a shift-symmetric quadratic function $\fm$ in \Eref{fs} which
remains invisible in $\Vhi$ while dominates the canonical
normalization of the Higgs-inflaton. Using $0.18~[0.19]\leq
m\leq10$ and confining $\rs$ to the range $(3.5\cdot10^{-3}-1/N)$
where $N=3$ [$N=2$] for $K=K_i$ [$K=K_{i+2}$] -- with $i=1,2$ --,
we achieved observational predictions which may be tested in the
near future and converge towards the ``sweet'' spot of the present
data. These solutions can be attained even with subplanckian
values of the inflaton requiring large $\cm$'s and without causing
any problem with the perturbative unitarity. It is gratifying,
finally, that our proposal remains intact from radiative
corrections, the Higgs-inflaton may assume ultimately its v.e.v
predicted by the gauge unification within MSSM, and the
inflationary dynamics can be studied analytically and rather
accurately.

As a last remark, we would like to point out that, although we
have restricted our discussion to the $\Ggut=\Gsm\times
U(1)_{B-L}$ gauge group, kinetically modified nMHI has a much
wider applicability. It can be realized, employing the same $W$
and $K$'s within other SUSY GUTs too based on a variety of gauge
groups -- such as the left-right \cite{fhi2}, the Pati-Salam
\cite{nmH}, or the flipped $SU(5)$ group \cite{fhi2} -- provided
that $\Phi$ and $\bar \Phi$ consist a conjugate pair of Higgs
superfields  so that they break $\Ggut$ and compose the gauge
invariant quantities $F_{\pm}$. Moreover, given that the term $\ld
M^2S/4$ of $W$ in \Eref{Win} plays no role during nMHI, our
scenario can be implemented by replacing it with $\kappa S^3$ and
identifying $\Phi$ and $\bar \Phi$ with the electroweak Higgs
doublets $H_u$ and $H_d$ of the next-to-MSSM \cite{linde1}. In
this case we have to modify the shift symmetry in \Eref{shift},
following the approach of \cref{shiftHI}, consider the soft SUSY
breaking terms to obtain the radiative breaking of $\Gsm$ and take
into account the renormalization-group running of the various
parameters from the inflationary up to the electroweak scale in
order to connect convincingly the high- with the low-energy
phenomenology. In all these cases, the inflationary predictions
are expected to be quite similar to the ones obtained here,
although the parameter space may be further restricted. The
analysis of the stability of the inflationary trajectory may be
also different, due to the different representations of $\Phi$ and
$\bar \Phi$. Since our main aim here is the demonstration of the
kinetical modification on the observables of nMHI, we opted to
utilize the simplest GUT embedding.



\def\ijmp#1#2#3{{\sl Int. Jour. Mod. Phys.}
{\bf #1},~#3~(#2)}
\def\plb#1#2#3{{\sl Phys. Lett. B }{\bf #1}, #3 (#2)}
\def\prl#1#2#3{{\sl Phys. Rev. Lett.}
{\bf #1},~#3~(#2)}
\def\rmp#1#2#3{{Rev. Mod. Phys.}
{\bf #1},~#3~(#2)}
\def\prep#1#2#3{{\sl Phys. Rep. }{\bf #1}, #3 (#2)}
\def\prd#1#2#3{{\sl Phys. Rev. D }{\bf #1}, #3 (#2)}
\def\npb#1#2#3{{\sl Nucl. Phys. }{\bf B#1}, #3 (#2)}
\def\npps#1#2#3{{Nucl. Phys. B (Proc. Sup.)}
{\bf #1},~#3~(#2)}
\def\mpl#1#2#3{{Mod. Phys. Lett.}
{\bf #1},~#3~(#2)}
\def\jetp#1#2#3{{JETP Lett. }{\bf #1}, #3 (#2)}
\def\app#1#2#3{{Acta Phys. Polon.}
{\bf #1},~#3~(#2)}
\def\ptp#1#2#3{{Prog. Theor. Phys.}
{\bf #1},~#3~(#2)}
\def\n#1#2#3{{Nature }{\bf #1},~#3~(#2)}
\def\apj#1#2#3{{Astrophys. J.}
{\bf #1},~#3~(#2)}
\def\mnras#1#2#3{{MNRAS }{\bf #1},~#3~(#2)}
\def\grg#1#2#3{{Gen. Rel. Grav.}
{\bf #1},~#3~(#2)}
\def\s#1#2#3{{Science }{\bf #1},~#3~(#2)}
\def\ibid#1#2#3{{\it ibid. }{\bf #1},~#3~(#2)}
\def\cpc#1#2#3{{Comput. Phys. Commun.}
{\bf #1},~#3~(#2)}
\def\astp#1#2#3{{Astropart. Phys.}
{\bf #1},~#3~(#2)}
\def\epjc#1#2#3{{Eur. Phys. J. C}
{\bf #1},~#3~(#2)}
\def\jhep#1#2#3{{\sl J. High Energy Phys.}
{\bf #1}, #3 (#2)}
\newcommand\jcap[3]{{\sl J.\ Cosmol.\ Astropart.\ Phys.\ }{\bf #1}, #3 (#2)}
\newcommand\njp[3]{{\sl New.\ J.\ Phys.\ }{\bf #1}, #3 (#2)}

\end{document}